\documentstyle{article}

\def\be{\begin{equation}}
\def\ee{\end{equation}}
\def\bea{\begin{eqnarray}}
\def\eea{\end{eqnarray}}
\begin{document}
\begin{titlepage}
\vspace*{10mm}
\begin{center}
{\large \bf Two dimensional fractional supersymmetric conformal field
theories and the two point functions}
\vskip 10mm
\centerline {\bf
 Fardin Kheirandish $^{a}$ \footnote {e-mail:fardin@iasbs.ac.ir} {\rm and}
 Mohammad Khorrami $^{a,b}$ \footnote {e-mail:mamwad@iasbs.ac.ir}}
 \vskip 1cm
{\it $^a$ Institute for Advanced Studies in Basic Sciences,}
\\ {\it P. O. Box 159, Zanjan 45195, Iran}\\
{\it $^b$ Institute for Studies in Theoretical physics and Mathematics,}\\
{\it P. O. Box 5531, Tehran 19395, Iran}\\
\end{center}
\vskip 2cm
\begin{abstract}
\noindent A general two dimensional fractional supersymmetric conformal
field theory is investigated. The structure of the symmetries of the theory
is studied. Then, applying the generators of the closed subalgebra generated
by $(L_{-1},L_{0},G_{-1/3})$ and
$(\bar{L}_{-1},\bar{L}_{0},\bar{G}_{-1/3})$, the two point functions of the
component fields of supermultiplets are calculated.
\end{abstract}
\end{titlepage}
\newpage
\section{Introduction}
2D conformally--invariant field theories have become the subject
of intense investigation in recent years, after the work of
Belavin, Polyakov, and Zamolodchikov [1]. One of the main reasons
for this, is that 2D conformal field theories describe the
critical behaviour of two dimensional statistical models [2--5].
Conformal field theory provides us with a simple and powerful
means of calculating the critical exponents, as well as, the
correlation functions of the theory at the critical point [1,6].
Another application of conformal field theories is in string
theories. Originally, string theory was formulated in flat twenty
six dimensional space--time for bosonic- and flat ten dimensional
space--time for supersymmetric-theories. It has been realized now
that the central part of string theory is a 2D
conformally--invariant field theory. It is also seen that tree
level string amplitudes may be expressed in terms of correlation
functions of the corresponding conformal field theory on the
plane, whereas string loop amplitudes may be expressed in terms of
correlation functions of the same conformal field theory on higher
genus Riemann surfaces [7--10].

Supersymmtry is a $Z_{2}$ extension of the Poincar\'e algebra [11,12]. But
this can be enlarged, to a superconformal algebra for example [13]. If the
dimension of the space--time is two, there are also fractional
supersymmetric extensions of the Poincar\'e and conformal algebra [14--17].
Fractional supersymmetry is a $Z_{n}$ extension of the Poincar\'e algebra.

In this paper, the special case $n=3$ is considered. So the
components of the superfield have grades 0, 1, and 2. The complex
plane is extended by introducing two independent paragrassmann
variables $\theta$ and $\bar{\theta}$, satisfying
$\theta^{3}={\bar{\theta}^{3}}=0$. One can develop an algebra, the
fractional $n=3$ algebra, based on these variables and their
derivatives [18--20]. In [21,22], this fractional supersymmetry
has been investigated by introducing a certain fractional
superconformal action. We don't consider any special action here.
What we do, is to use only the structure of fractional
superconformal symmetry in order to obtain general restrictions on
the two--point functions. The scheme of the paper is the
following. In section 2, infinitesimal superconformal
transformations are defined. In section 3, the generators of these
transformation and their algebra are investigated. In section 4,
the two--point functions of such theories are obtained. Finally,
section 5 contains the concluding remarks.
\section{Infinitesimal superconformal transformations}
Consider a paragrassmann variable $\theta$, satisfying
\be
\theta^3=0. \ee A function of a complex variable $z$, and this
paragrassmann variable, will be of the form
\be
f(z,\theta)=f_0(z)+\theta f_1(z)+\theta^2 f_2(z).
\ee
Now define the covariant derivatives as [16,18,23,24]
\be\label{1}
D:=\partial_{\theta}-q\theta^{2}\partial_{z},
\ee
where $q$ is one of the third roots of unity, not equal to one, and
$\partial_\theta$ satisfies
\be
\partial_\theta \theta=1+q\theta\partial_\theta.
\ee
An infinitesimal transformation
\bea\label{3}
z'&=&z+\omega_{0}(z)+\theta\omega_{1}(z)+\theta^{2}\omega_{2}(z)\cr
\theta'&=&\theta+\epsilon_{0}(z)+\theta\epsilon_{1}(z)+
\theta^{2}\epsilon_{2}(z)
\eea
is called superconformal if
\be\label{4}
D=(D\theta')D'
\ee
where
\be
D'=\partial_{\theta'}-q\theta'^{2}\partial_{z'} \ee From these, it
is found that an infinitesimal superconformal transformation is of
the form \bea\label{6} \theta'&=&\theta+\epsilon_{0}(z)+{1\over
3}\theta\omega'_{0}(z)+ \theta^{2}\epsilon_{2}(z),\cr
z'&=&z+\omega_{0}(z)-q\theta^{2}\epsilon_{0}(z), \eea where
$\omega'_0(z):=\partial_z\omega_0(z)$. It is also seen that the
following commutative relations hold.
 \bea\label{7}
\epsilon_{2}\theta&=&q\theta\epsilon_{2},\cr
\epsilon_{0}\theta&=&q\theta\epsilon_{0}. \eea

One can extend these naturally to functions of $z$ and $\bar z$,
and $\theta$ and $\bar\theta$ (full functions instead of chiral
ones). It is sufficient to define a covariant derivative for the
pair $(\bar z,\bar\theta)$, the analogue of (3), and extend the
transformations (5), so that there are similar transformations for
$(\bar z,\bar\theta)$ as well. Then, defining a superconformal
transformation as one satisfying (6) and its analogue for $(\bar
z,\bar\theta)$, one obtains, in addition to (8) and (9), similar
expressions where $(z,\theta,\omega,\epsilon)$ are simply replaced
by $(\bar z,\bar\theta,\bar\omega,\bar\epsilon)$. So, the
superconformal transformations consist of two distinct class of
transformations, the holomorphic and the antiholomorphic, that do
not talk to each other.
\section{Generators of superconformal field theory }
The (chiral) superfield $\phi(\theta, z)$ with the expansion [24]
\be
\phi(\theta,z)=\varphi_{0}(z)+\theta\varphi_{1}(z)+
\theta^{2}\varphi_{2}(z)
\ee
is a super-primaryfield of weight $\Delta$ if it transforms under a
superconformal transformations as
\be
\phi(\theta, z)\mapsto(D\theta')^{3\Delta}\phi(\theta',z')
\ee
One can write this as
\be
\phi(\theta, z)\mapsto [1+\tilde T(\omega_0)+\tilde S(\epsilon_0)+
\tilde H(\epsilon_2)]\phi(\theta,z),
\ee
to arrive at [18]
\bea
\tilde{T}(\omega_{0})&=&\omega_{0}\partial_{z}+\left(\Delta
+{\Lambda\over 3}\right)\omega'_{0},\cr
\tilde{S}(\epsilon_{0})&=&\epsilon_{0}(\delta_{\theta}-q^{2}\theta^{2}
\partial_{z})-3\Delta q^{2}\epsilon'_{0}\theta^{2},\cr
\tilde{H}(\epsilon_2)&=&q\epsilon_{2}(\theta^{2}\partial_{\theta}-
\Delta\theta).
\eea
Here $\Lambda$ and $\delta_\theta$ are operators satisfying
\be
[\Lambda,\theta]=1,
\ee
and
\be
\delta_\theta\theta=q^{-1}\theta\delta_\theta+1,
\ee
respectively.

One can now define the generators
 \bea l_n&:=&\tilde
T(z^{n+1}),\cr g_r&:=&\tilde S(z^{r+1/3}), \eea where $n$ and
$r+1/3$ are integers. The generators of superconformal
transformations are defined through \bea\label{17}
[L_n,\phi(\theta, z)]&:=&l_n\phi,\cr [G_r,\phi(\theta
,z)]&:=&g_r\phi. \eea One can check that, apart from a possible
central extension, these generators satisfy the following
relations. \be\label{30} [L_n,L_m]=(n-m)L_{n+m}, \ee
\be
[L_n,G_r]=\left({n\over 3}-r\right)G_{n+r},
\ee
and
\be
G_rG_sG_t+\hbox{five other permutations of the
indices}=6L_{r+s+t}. \ee This algebra, which contains the Virasoro
algebra (18) as a subalgebra, has nontrivial central extensions.
It is shown [16,25] that there is only one subalgera (containing
$G$-generators as well as $L_n$'s), the central extension for
which is trivial. This algebra is the one generated by
$\{L_{-1},L_0,G_{-1/3}\}$. Note that if one excludes $G_r$'s,
there exists another subalgebra generated by
$\{L_{-1},L_0,L_{1}\}$, the central extension of which is trivial.
Also note that we have not included the generators corresponding
to $\tilde H$ in the algebra. The reason is that there is no
closed subalgebra, with a trivial central extension, containing
these generators [16,25].

Now we have the effects of $L_n$'s and $G_r$'s on the superfield.
It is not difficult to obtain their effect on the component
fields. For $L_n$'s, the first equation of (17) leads directly to
\be\label{21} [L_n,\phi_k(z)]=z^{n+1}\partial_z\phi_k+(n+1)z^n
\left(\Delta+{k\over 3}\right)\phi_k. \ee This shows that the
component field $\phi_k$ is simply a primary field with the weight
$\Delta +k/3$. One can also write (21) in terms of
operator--product expansion:
\be
{\cal R}[T(w)\phi_k(z)]\sim{{\partial\phi_k(z)}\over{w-z}}+
{{(\Delta +k/3)\phi_k(z)}\over{(w-z)^2}}, \ee where ${\cal R}$
denotes the radial ordering and $T(z)$ is the holomorphic part of
the energy--momentum tensor:
\be
T(z)=\sum_n{{L_n}\over{z^{n+2}}}.
\ee

For $G_r$'s, a little more care is needed. One defines a $\chi$--commutator
as [26]
\be
[A,B]_\chi:=AB-\chi BA.
\ee
It is easy to see that
\be
[A,BB']_{\chi\chi'}=[A,B]_\chi B'+\chi B[A, B']_{\chi'}.
\ee
Now, if we use
\be
[G,\theta]_q=0,
\ee
then the second equation of (\ref{17}) leads to
\bea\label{27}
[G_r,\phi_0(z)]&=&z^{r+1/3}\phi_1,\cr
[G_r,\phi_1(z)]_{q^{-1}}&=&-z^{r+1/3}\phi_2,\cr
[G_r,\phi_2(z)]_{q^{-2}}&=&-\left[z^{r+1/3}
\partial_z\phi_0+\left(r+{1\over 3}\right)z^{r-2/3}(3\Delta)\phi_0\right].
\eea This can also be written in terms of the operator--product
expansion. To do this, however, one should first define a proper
radial ordering for the supersymmetry generator and the component
fields. Defining
\be
{\cal R}[S(w)\phi_k(z)]:=\cases{S(w)\phi_k(z),&$|w|>|z|$\cr
                          q^{-k}\phi_k(z)S(w),&$|w|<|z|$\cr}
\ee
where
\be
S(z):=\sum_r{{G_r}\over{z^{r+4/3}}}, \ee one is led to \bea {\cal
R}[S(w)\phi_0(z)]&\sim&{{\phi_1(z)}\over{w-z}},\cr {\cal
R}[S(w)\phi_1(z)]&\sim&-{{\phi_2(z)}\over{w-z}},\cr {\cal
R}[S(w)\phi_2(z)]&\sim&-{{\partial\phi_0(z)}\over{w-z}}-
{{3\Delta\phi_0(z)}\over{(w-z)^2}}. \eea

What we really use to restrict the correlation functions is that
part of the algebra the central extension of which is trivial,
that is, the algebra generated by $\{ L_{-1},L_0,G_{-1/3}\}$.

\section{Two--point functions}
The two--point functions should be invariant under the action of
the subalgebra generated by $\{ L_{-1},L_0,G_{-1/3}\}$. This means
\bea\label{31} \langle 0|[L_{-1},\phi_k\phi'_{k'}]|0\rangle&=&0,\\
\label{32} \langle 0|[L_0,\phi_k\phi'_{k'}]|0\rangle&=&0,\\
\label{33} \langle
0|[G_{-1/3},\phi_k\phi'_{k'}]_{q^{-k-k'}}|0\rangle&=&0. \eea Here
we have used the shorthand notation $\phi_k=\phi_k(z)$ and
$\phi'_{k'}=\phi'_{k'}(z')$. $\phi$ and $\phi'$ are primary
superfields of weight $\Delta$ and $\Delta'$, respectively.
Equations (31) and (32) imply that
\be
\langle\phi_k\phi'_{k'}\rangle ={{A_{k,k'}}\over
{(z-z')^{\Delta +\Delta'+(k+k')/3}}}.
\ee
This is simply due to the fact that $\phi_k$ and $\phi'_{k'}$ are primary
fields of the weight $\Delta +k/3$ and $\Delta' +k'/3$, respectively. Note
that it is not required that these weights be equal to each other, since we
have not included $L_1$ in the subalgebra.

(33) relates $A_{k_1,k'_1}$ with $A_{k_2,k'_2}$, if
\be
k_1+k'_1-(k_2+k'_2)=0,\qquad\hbox{mod 3}. \ee Therefore, there
remains 3 independent constants in the 9 correlation functions.
Correlation functions of grade 0: \bea\label{36}
\langle\phi_0\phi'_0\rangle&:=&A_0f_{0,0}=
A_0{1\over{(z-z')^{\Delta +\Delta'}}},\cr
\langle\phi_1\phi'_2\rangle&:=&A_0f_{1,2}=
A_0{{\Delta+\Delta'}\over{(z-z')^{\Delta +\Delta'+1}}},\cr
\langle\phi_2\phi'_1\rangle&:=&A_0f_{2,1}=
A_0{{-q^2(\Delta+\Delta')}\over{(z-z')^{\Delta +\Delta'+1}}}, \eea
those of grade one: \bea\label{37}
\langle\phi_0\phi'_1\rangle&:=&A_1f_{0,1}=
A_1{1\over{(z-z')^{\Delta +\Delta'+1/3}}},\cr
\langle\phi_1\phi'_0\rangle&:=&A_1f_{1,0}=
A_1{{-1}\over{(z-z')^{\Delta +\Delta'+1/3}}},\cr
\langle\phi_2\phi'_2\rangle&:=&A_1f_{2,2}=
A_1{{q^2(\Delta+\Delta'+1/3)}\over{(z-z')^{\Delta +\Delta'+4/3}}},
\eea and those of grade two: \bea\label{38}
\langle\phi_0\phi'_2\rangle&:=&A_2f_{0,2}=
A_2{1\over{(z-z')^{\Delta +\Delta'+2/3}}},\cr
\langle\phi_1\phi'_1\rangle&:=&A_2f_{1,1}=
A_2{1\over{(z-z')^{\Delta +\Delta'+2/3}}},\cr
\langle\phi_2\phi'_0\rangle&:=&A_2f_{2,0}=
A_2{{q^2}\over{(z-z')^{\Delta +\Delta'+2/3}}}. \eea

It is easy to check that adding $L_1$ to the subalgebra
trivializes the correlation functions. The reason is that in all
of the correlation sets (of the same grade) there exists at least
one correlation function the weights of its primary fields are not
equal to each other, regardless of the values of $\Delta$ and
$\Delta'$. In fact, acting on $\langle\phi_k\phi'_{k'}\rangle$ by
$G_{-1/3}$ relates $\langle\phi_{k+1}\phi'_{k'}\rangle$ to
$\langle\phi_k\phi'_{k'+1}\rangle$. And it is impossible that both
\be
\Delta+1/3=\Delta',\quad\hbox{mod 1}
\ee
and
\be
\Delta=\Delta'+1/3,\quad\hbox{mod 1}
\ee
hold. This shows that either
$\langle\phi_{k+1}\phi'_{k'}\rangle$ or $\langle\phi_k\phi'_{k'+1}\rangle$
are zero, and so the other should be zero as well.

So far, everything has been calculated for the chiral fields. Let
us generalize this no the full fields. The generalization is not
difficult. One introduces component fields $\phi_{k\bar k}(z,\bar
z)$ which have the following properties. The action of the
energy--momentum generators on these component fields is \bea
[L_n,\phi_{k\bar k}(z,\bar z)]&=&z^{n+1}\partial_z\phi_{k\bar k}+
(n+1)z^n\left(\Delta+{k\over 3}\right)\phi_{k\bar k},\cr [\bar
L_n,\phi_{k\bar k}(z,\bar z)]&=&\bar z^{n+1}\partial_{\bar z}
\phi_{k\bar k}+(n+1)\bar z^n\left(\bar\Delta+{{\bar k}\over
3}\right) \phi_{k\bar k}. \eea While there are generators of the
supersymmetry which act like \bea [G_r,\phi_{0\bar k}(z,\bar
z)]_{q^{-\bar k}}&=&z^{r+1/3}\phi_{1\bar k},\cr [G_r,\phi_{1\bar
k}(z,\bar z)]_{q^{-1-\bar k}}&=&-z^{r+1/3}\phi_{2\bar k}, \cr
[G_r,\phi_{2\bar k}(z,\bar z)]_{q^{-2-\bar k}}&=&-\left[z^{r+1/3}
\partial_z\phi_{0\bar k}+\left(r+{1\over 3}\right)z^{r-2/3}(3\Delta)
\phi_{0\bar k}\right],
\eea
and
\bea
[\bar G_r,\phi_{k0}(z,\bar z)]_{q^{-k}}&=&\bar z^{r+1/3}\phi_{k1},\cr
[\bar G_r,\phi_{k1}(z,\bar z)]_{q^{-k-1}}&=&-\bar z^{r+1/3}\phi_{k2},\cr
[\bar G_r,\phi_{k2}(z,\bar z)]_{q^{-k-2}}&=&-\left[\bar z^{r+1/3}
\partial_{\bar z}\phi_{k0}+\left(r+{1\over 3}\right)\bar z^{r-2/3}
(3\bar\Delta)\phi_{k0}\right].
\eea

The equations to be satisfied by the two--point functions are
\bea
\langle 0|[L_{-1}(\bar L_{-1}),\phi_{k\bar k}
\phi'_{k'\bar k'}]|0\rangle=0,\cr
\langle 0|[L_0(\bar L_0),\phi_{k\bar k}
\phi'_{k'\bar k'}]|0\rangle=0,\cr
\langle 0|[G_{-1/3}(\bar G_{-1/3}),\phi_{k\bar k}
\phi'_{k'\bar k'}]_{q^{-k-\bar k-k'-\bar k'}}|0\rangle=0.
\eea
Once again, the first two sets of equations simply imply that
\be
\langle\phi_{k\bar k}\phi'_{k'\bar k'}\rangle={{A_{k\bar k,k'\bar
k'}}\over {(z-z')^{\Delta+\Delta'+(k+k')/3}(\bar z-\bar z')^{\bar
\Delta+\bar \Delta'+(\bar k+\bar k')/3}}}. \ee The third set of
equations relate $A_{k_1\bar k_1,k'_1\bar k'_1}$ to $A_{k_2\bar
k_2,k'_2\bar k'_2}$, provided
\be
k_1+k'_1-(k_2+k'_2)=\bar k_1+\bar k'_1-(\bar k_2+\bar
k'_2)=0,\qquad \hbox{mod 3}. \ee So, there remains nine arbitrary
constants in these correlation functions. One can write the
correlations in terms of the correlations of the chiral fields. To
see this, notice that
\be
\langle 0|[G_{-1/3},\phi_{k\bar k}\phi'_{k'\bar k'}]_{q^{-k-\bar
k}} |0\rangle=O_k\langle\phi_{(k+1)\bar k}\phi'_{k'\bar k'}\rangle
+ q^{-k-\bar k}O'_{k'}\langle\phi_{k\bar k}\phi'_{(k'+1)\bar
k'}\rangle. \ee Here $O_k$ and $O'_{k'}$ are coefficients (or
derivatives) which depend only on $k$ and $k'$, respectively. The
right--hand should of course be zero, due to the supersymmetry of
the theory. The corresponding equation for chiral correlations
differs from this, only in the coefficient $q^{-k-\bar k}$. For
the chiral correlations, this coefficient is simply $q^{-k}$. But
it is easy to check that the set $q^{k\bar k}\langle\phi_{k\bar
k}\phi'_{k'\bar k'}\rangle$ satisfies the same equations of the
chiral correlations. It is also easy to see that this same set
satisfy the same equations of the antichiral correlations as well.
So, the general solution for the full correlations is
\be
\langle\phi_{k\bar k}\phi'_{k'\bar k'}\rangle=A_{K\bar K}q^{-k\bar
k} f_{k,k'}(z-z')\bar f_{\bar k,\bar k'}(\bar z-\bar z'), \ee
where $f_{k,k'}$'s are defined through (36)--(38), $\bar f_{\bar
k,\bar k'}$'s are the same as these with $\Delta\to\bar\Delta$ and
$\Delta'\to\bar\Delta'$,
\be
K=k+k',\qquad\hbox{mod 3},
\ee
and
\be
\bar K=\bar k+\bar k',\qquad\hbox{mod 3}.
\ee
\section{Concluding remarks}
We considered fractional superconformal field theories only
through the general symmetries of such theories. For the special
case of three component fields in the superfield, the two--point
functions were obtained. In fractional superconformal field
theories, one cannot impose the symmetries generated by $L_1$ and
$\bar L_1$ as non anomalous symmetries. Otherwise, the whole
theory becomes trivial. This is due to the fact that the
fractional supersymmetry generator changes the weight of the
fields by $1/3$ (or $1/f$ if there are $f$ component fields in the
superfield). So, two two--point functions will be related to each
other, and the difference between the weights of the fields of one
correlation is $2/f$ greater than the difference between the
weights of the fields of the other correlation. As the symmetry
generated by $L_1$ requires the weights of the fields in nonzero
two--point functions be equal to each other, at least one of these
correlations will be zero, and this forces the other to be zero as
well. One can also see that the theory becomes trivial, through
the fact that adding $L_1$ to the subalgebra generated by $\{
L_{-1},L_0,G_{-1/3}\}$ brings all of the generators of the
superconformal theory to the algebra.

As $L_1$ is not in the non anomalous symmetry generators, the
component fields don't enjoy the full conformal symmetry (that is
the $sl(2)$ symmetry generated by $\{ L_{-1},L_0,L_1\}$). So, the
three--point functions cannot be determined up to only constants;
there remains unknown functions in them. In fact one has lost the
special conformal transformations as non anomalous symmetries, and
the symmetries of the theory, apart from the supersymmetry, are
just the space--time symmetries plus dilation.


\newpage
\begin{thebibliography}{99}
\bibitem{1} A. A. Belavin, A. M. Polyakov, A. B. Zamolodchikov,
Nucl. Phys. B241 (1984) 333
\bibitem{2} Manuel Asorey, Fortschr. Phys. 40 (1992) 3, 273
\bibitem{3} P. Christe and M. Henkel, Introduction to Conformal Invariance
and it's Applications to Critical Phenomena, Springer Lecture
Notes in Physics, Vol. m 16 (1993)
\bibitem{4} John L. Cardy, Les Houches, session XLIX, 1988
            /fields, strings and critical phenomena, ed. E. Brezin
            and J. Zinn-Justin (North-Holland, Amsterdam, 1990)
\bibitem{5} Philippe Di Francesco, Pierre Mathieu and David Senecal,
 Conformal Field Theory, Graduate Texts in Contemporary Physics,
 1997 Springer-Verlag, New York, Inc.
\bibitem{6} Daniel Friedan, Zongan Qiu, and Stephen Shenker, Phys.
Rev. lett. 52, (1984) 1575
\bibitem{7} P. Ginsparg, Les Houches, session XLIX, 1988
            /fields, strings and critical phenomena, ed. E. Brezin
            and J. Zinn-Justin (North-Holland, Amsterdam, 1990)
\bibitem{8} Lance J. Dixon, Introduction to conformal field theory
and string theory, slac-pub-5149, 1989
\bibitem{9} Elias Kiritsis, Introduction to superstring theory, preprint CERN-TH/97-218, hep-th/9709062
\bibitem{10} Joseph Polchinski, What is string theory, preprint NSF-ITP-94-97, hep-th/9411028
\bibitem{11} S. Coleman and J. Mandula, Phys. Rev 159 (1967) 1251
\bibitem{12} R. Haag, J. T. Lopuszanski and M. F. Sohnius, Nucl.
Phys. B88 (1975) 257
\bibitem{13} D. Friedan, Z. Qiu and S. Shenker, Phys. lett. 151B
(1985) 37
\bibitem{14} C. Ahn, D. Bernard and A. Leclair, Nucl. Phys. B346
(1990) 409
\bibitem{15} J. A. de Azcarraga and A. J. Macfarlane, J. Math.
Phys. 37 (1996) 111
\bibitem{16} S. Durand, Phys. Lett B 312 (1993) 115
\bibitem{17} N. Fleury and M. Raush de Traubenberg, Mod. Phys.
Lett, A11 (1996) 899
\bibitem{18} A. T. Philippov, A. P. Isaev and A. B. Kurdikov,
Paragrassmann extensions of the Virasoro algebra, preprint
hep-th/9212157
\bibitem{19} S. Durand, Mod. Phys. Lett A7 (1992) 2905
\bibitem{20} A. T. Filippov and A. B. Kurdikov, Paragrassmann algebras
 with many variables, preprint hep-th/9312081,
              A. P. Isaev, Paragrassmann integral, discrete systems
               and quantum groups, preprint q-alg/9609030
\bibitem{21} A. Perez, M. Rausch de Traubenberg, and P. simon,
Nucl. Phys. B 482 (1996) 325
\bibitem{22} M. Rausch de Traubenberg and P. Simon, 2D-Fractional
 supersymmetry and conformal field theory for alternative statistics, preprint hep-th/9606188
\bibitem{23} S. Durand, Mod. Phys. Lett A8(1993) 2323
\bibitem{24} E H Saidi, M B sedra and Zarouaoui, Class. Quantum.
Grav. 12 (1995) 1567
\bibitem{25} S. Durand and L. Vinet, Mod. Phys. Lett A4, 2519
(1989)
\bibitem{26} R. S. Dunne, A. J. Macfarlane, J. A. Azcarraga et J.
C. Perez Bueno, Int. J. Mod. Phys. A12 (1997) 3275
\bibitem{27} Zongan Qiu, Nucl. Phys. B 270 [Fs16] (1986) 205
\end{thebibliography}
\end{document}